\documentclass[twocolumn,nodate,preprintnumbers,nofootinbib,amsmath,amssymb,aps,prd]{revtex4}

\usepackage{setspace}
\usepackage{amssymb}
\usepackage{graphicx,wrapfig}
\usepackage{enumerate} 
\usepackage{color}
\usepackage{mathrsfs}
\usepackage{subfigure}
\usepackage[mathscr]{euscript}
\definecolor{mg}{rgb}{0.0, 0.5, 0.0}

\def\be{\nopagebreak[3]\begin{equation}}
\def\ee{\end{equation}}
\def\ba{\nopagebreak[3]\begin{eqnarray}}
\def\ea{\end{eqnarray}}

\begin{document}

\title{What can Black Holes teach us about the IR and UV?}
\author{Basem Kamal El-Menoufi}
\email{basem.el-menoufi@manchester.ac.uk}
\affiliation{
 Department of Physics and Astronomy, University of Sussex, Brighton, BN1 9QH, United
Kingdom
}
\author{Sonali Mohapatra}
 \email{s.mohapatra@sussex.ac.uk}
\affiliation{
 Department of Physics and Astronomy, University of Sussex, Brighton, BN1 9QH, United
Kingdom
}

\begin{abstract}
Combining insights from both the effective field theory of quantum gravity and black hole thermodynamics, we derive two novel consistency relations to be satisfied by any quantum theory of gravity. First, we show that a particular combination of the number of massless (light) fields in the theory must take integer values. Second, we show that, once the massless spectrum is fixed, the Wilson coefficient of the Kretschmann scalar in the low-energy effective theory is fully determined by the logarithm of a single natural number.
\end{abstract}

\maketitle

\emph{Introduction -} The link between black holes and their intrinsic thermodynamical behavior is perhaps {\em the} key to a consistent theory of quantum gravity. The underlying quantum degrees of freedom of a black hole must necessarily account for its entropy \cite{Strominger:2009aj, Rovelli:1996dv, tHooft:1993dmi, Page:2004xp}. Although currently we are far from describing black hole ``micro-states", powerful insights were revealed relying on effective field theory lore. Indeed, Hawking radiation was discovered by considering the quantum dynamics of matter fields in a fixed black hole geometry \cite{Hawking:1974sw}.

Donoghue \cite{Donoghue:1994dn} demonstrated that quantum gravity, at distances large compared to the Planck length, is well described by an effective field theory (EFT). It is remarkable that gravity lends itself naturally to the EFT framework. All the unknown physics coming from the ultraviolet (UV) is encoded solely in the Wilson coefficients of the most general diffeomorphism-invariant action. Similar to any UV-sensitive quantity in quantum field theory, the Wilson coefficients can only be determined empirically. More importantly, long-distance quantum effects furnish a set of reliable and parameter-free predictions of the EFT, as they emerge from the low-energy portion of loops containing massless (light) degrees of freedom. 

Recently, some work has been done to adapt and utilize the EFT framework to study quantum aspects of black hole thermodynamics in the context of Euclidean quantum gravity \cite{El-Menoufi:2015cqw, El-Menoufi:2017kew}. In particular, it was shown that the long-distance contribution to the partition function is captured by covariant non-local operators. This is an interesting development which, in particular, allows us to quantify a set of quantum corrections to the various thermodynamic relations governing black holes. Notably, it was shown in \cite{El-Menoufi:2015cqw, El-Menoufi:2017kew} that the non-local operators are responsible for generating the logarithmic correction to the Bekenstein-Hawking entropy of Schwarzschild black hole. 

The advent of gravitational wave observations \cite{Abbott:2016blz} has revived interest in using experimental data to test black hole thermodynamics \cite{foit, Bianchi:2018ula}. In this paper we aim to complement this endeavor, albeit by following a different approach. In particular, we utilize the structure of the logarithmic contribution to the entropy, obtained through the EFT framework, to derive two consistency relations that hold for any theory (model) of quantum gravity. This is primarily achieved by invoking Bekenstein's conjecture \cite{Bekenstein:1974jk} that the area of Schwarzschild black hole has a discrete spectrum in any quantum theory of gravity. In essence, the universality of the EFT and black hole thermodynamics are the two main motivations behind such relations. 

The first relation sets a constraint on the number of massless (light) fields coupled to gravity. To an effective field theorist, this is indeed remarkable because gravity can generally couple to any number of massless fields. The second relation constrains the Wilson coefficient of the Kretschmann scalar, measured at an arbitrary scale, to be determined in terms of a single natural number. This is striking given that experimental bounds are exceedingly weak on the Wilson coefficients of quadratic gravity \cite{Hoyle:2004cw}. We will now briefly review the basic ingredients necessary for the derivation of the consistency conditions which will follow. 
 \\ \\
\emph{Black Hole Area Quantization -}
The proposal that the area of a black hole is quantized relies on the observation that the area appears to behave {\em classically} as an adiabatic invariant. The initial evidence for this conjecture came from the work of Christodoulou and Ruffini \cite{Christodoulou:1970wf, Ruf}. In particular, they showed that the area of a non-extremal black hole does not change in the process of absorbing a classical point particle, if the capture takes place at the turning point of the particle's orbit. The implied reversibility of this process hints towards the adiabatic invariance of the horizon area \cite{Bekenstein:1997bt}. Ehrenfest's hypothesis \cite{Bekenstein:1997bt, Erhenfest} states that any physical quantity, which is classically an adiabatic invariant, is quantized in the quantum theory. Following this line of reasoning, one is presented with Bekenstein's quantization conjecture: The area of a black hole is quantized in any theory of quantum gravity. 

Now the natural question is, how does this quantized area spectrum look like?
Ascribing quantum mechanical uncertainty to the particle captured by the black hole, it is not difficult to show that there exists a {\em minimal} increase in the horizon area \cite{Bekenstein:1973ur, Bekenstein:1997bt}. This motivated the following spectrum \cite{Bekenstein:1997bt, Bekenstein:1995ju},
\begin{align}\label{BMquant}
\mathcal{A}_{n} = \gamma_0 l_p^2 n, \quad n = 1,2,3...  \:,
\end{align}
where $\gamma_0$ is some number that will be discussed later. For some exhaustive reviews on this topic, the reader is directed towards \cite{Bekenstein:1997bt, Jacobson:2018nnf}. The discrete nature of the area spectrum has also been observed in quantum gravity approaches, such as string theory \cite{Kogan:1994fg}, and loop quantum gravity \cite{Rovelli:96, Krasnov}. Nevertheless, there is no consensus regarding the uniformity of the spacing. In Loop Quantum Gravity \cite{Corichi:2006wn}, in particular, the area shows a highly non-uniform spectrum. In this paper, remaining agnostic to the spacing, we use a generalized quantization rule for the area as our starting point \cite{Hod:2004di}. This will allow us not to dwell on any particular model of quantum gravity.
\\ \\
\emph{EFT and black hole Thermodynamics -} Hawking and Gibbons pioneered a consistent approach to study the thermal properties of black holes \cite{Gib1977}. For a gravitational system at finite temperature, the partition function of the canonical ensemble reads
\begin{align}\label{Zdef}
Z(\beta) = \int \mathcal{D} \Psi \mathcal{D} g \, e^{-\mathcal{S}_E - \mathcal{S}_{\partial}} \ \ .
\end{align}
where $\Psi$ stands for all matter fields coupled to gravity, $\mathcal{S}_E$ is the Euclidean action and $\mathcal{S}_{\partial}$ denotes the Hawking-Gibbons-York boundary action \cite{Gross:1982cv, York:1972sj}. As is customary from finite temperature field theory, the integral extends over (anti)-periodic field configurations for bosons (fermions). For gravity the prescription is to sum over positive-definite metrics with a fixed induced metric on the boundary. In the canonical ensemble, the boundary geometry is flat space on $\mathbb{S}_1 \times \mathbb{R}^3$ with the circumference of the time circle given by $\beta$. 

In a semi-classical evaluation of the partition function, the Euclidean section of the black hole appears as a saddle point. In \cite{El-Menoufi:2015cqw}, it was shown how to apply the techniques of the EFT of quantum gravity to compute the partition function, Eq.~(\ref{Zdef}). As we alluded to in the introduction, the partition function contains non-local operators which encapsulate the long distance dynamics. The result simplifies if the background geometry is a Kerr-Schild spacetime \cite{El-Menoufi:2015cqw}, which is the case for Schwarzschild black hole. At one loop, or more precisely at next-to-leading order in the EFT expansion, the partition function of Schwarzschild black hole in 4D is obtained by a Wick rotation of the effective action \cite{El-Menoufi:2017kew}
\begin{align}\label{partitionfunc}
\ln Z = \Gamma_{\text{local}}[\bar{g}] + \Gamma_{\text{ln}}[\bar{g}]  - \mathcal{S}_{\partial} \ \ ,
\end{align} 
where $\bar{g}$ denotes the background Kerr-Schild spacetime. First, we have
\begin{align}\label{localeff}
\Gamma_{\text{local}}[\bar{g}; \mu] &= \int d^4 x \Bigl[ \frac{M_P^2}{2} R +c_1^r(\mu)R^2   \nonumber \\
\qquad &+ c_2^r(\mu) R_{\mu\nu} R^{\mu\nu} + c_3^r(\mu) R_{\mu\nu\alpha\beta}R^{\mu\nu\alpha\beta} \nonumber \\
\qquad  &+  c^r_4(\mu) \Delta R + \mathcal{O}(R^3)  \Bigr] \ \ .
\end{align}
In the above, $\mu$ is the scale of dimensional regularization, $\Delta$ is the $4\text{D}$ flat Laplacian on ${\rm I\!R}^3 \times S^1$, and let us note that the $c^r_i$ are renormalized Wilson coefficients. Second, the non-local portion reads
\begin{align}\label{logspecies}
\nonumber
\Gamma_{\text{ln}}[\bar{g}] &= - \int d^4x \, \bigg[\alpha R \ln\left(\frac{-\Delta}{\mu^2}\right) R + \beta R_{\mu\nu} \ln\left(\frac{-\Delta}{\mu^2}\right) R^{\mu\nu} \\
&+ \gamma R_{\mu\nu\alpha\beta} \ln\left(\frac{-\Delta}{\mu^2}\right) R^{\mu\nu\alpha\beta} + \Theta \ln\left(\frac{-\Delta}{\mu^2}\right) \Delta R \bigg] \ \ ,
\end{align}
where the coefficients are finite numbers borne out of the calculation and depend on the spin of the massless field \cite{El-Menoufi:2015cqw}. The action in Eq.~(\ref{logspecies}) is expressed in quasi-local form, and we truncated the partition function at second order in the curvature expansion. The possible effects of the higher curvature operators on the thermodynamics were thoroughly discussed in \cite{El-Menoufi:2017kew}. Of most importance to our analysis is the invariance of the partition function under the renormalization group flow. Explicitly, the beta function of the Kretchman scalar coefficient is
\begin{equation}
\beta_{c_3} = - 2 \gamma \ \ .
\end{equation}
The logarithmic {\em form factor} in Eq.~(\ref{logspecies}), $\ln{\left(-\Delta/\mu^2\right)}$ at first glance, is a very complicated object. Indeed, the form factor is an integration kernel that must be evaluated in position space and takes the form of a distribution \cite{El-Menoufi:2017kew}
\begin{align}
\mathcal{L}(\vec{x}- \vec{x}') &=  -\frac{1}{2\pi} \lim_{\epsilon \rightarrow 0} \Big[ \mathcal{P} \left( \frac{1}{|\vec{x}- \vec{x}'|^3}\right) \nonumber \\
\qquad&+ 4\pi \left(\ln{(\mu \epsilon)} + \gamma_{E} - 1 \right)\delta^3(\vec{x}) \Big] \ \ ,
\end{align}
where $\mathcal{P}$ stands for principal value. With the kernel in hand, a direct evaluation of Eq.~(\ref{logspecies}) is possible \cite{El-Menoufi:2017kew} 
\begin{align}\label{poartfun}
\ln Z(\beta) = - \frac{\beta^2}{16\pi G} +  \bigg[64\pi^2 c^r_3(\mu) + 2 \, \Xi \ln(\mu \beta)\bigg] \ \ ,
\end{align} 
where $\beta = 8\pi G M$ and $\Xi$ counts the number of light fields minimally coupled to gravity
\begin{align}\label{sumparticles}
\Xi = \frac{1}{180} \left(2 N_S + \frac72 N_F - 26 N_V + 424 \right).
\end{align}
In the above, $N_S, N_F$ and $N_V$ are the number of scalars, Weyl-fermions and vectors in our theory, and the number $424$ is the contribution due to pure gravity. Using the partition function one can immediately compute the logarithmic correction to Bekenstein-Hawking entropy \cite{El-Menoufi:2015cqw}
\begin{equation}\label{Area-Entropy}
S_{\text{bh}} = \frac{\mathcal{A}}{4G} +  \left( 64\pi^2c_3^r(\mu) + \Xi \ln{(\mu^2 \mathcal{A})} \right) \ \ ,
\end{equation}
where $\mathcal{A} = 16\pi(GM)^2$ is the horizon area of a Schwarzschild black hole\footnote{The logarithmic correction to the entropy has been the subject of continuous interest, see \cite{Fur1995, Sen2012, Car2000, Sun2001, Ban2009, Aros, Cai2010, Kau2000}.}. The above structure reveals the power of the EFT, in particular, there is a clear separation between the short distance and long-distance dynamics. All the unknown ultraviolet physics is encoded in the renormalized Wilson coefficient, while the logarithm of the area emerges from the infrared structure of the theory. 

Furthermore, we can manifest the invariance of Eq.~(\ref{Area-Entropy}) under the renormalization group by using dimensional transmutation. The constant $c^r_3(\mu)$ is traded off for a dimensionful quantity, $\mathcal{A}_{\text{QG}}$, as follows
\begin{align}\label{dimtrans}
c^r_3(\mu)  = - \frac{\Xi}{64\pi^2} \ln \left( \mu^2 \mathcal{\mathcal{A}_{\text{QG}}}\right) \ \ ,
\end{align}
and thus Eq.~(\ref{Area-Entropy}) takes the elegant form
\begin{align}\label{logentropy}
S_{\text{bh}} = S_{\text{BH}} + \Xi \, \ln \left(\frac{\mathcal{A}}{\mathcal{A}_{\text{QG}}}\right) \ \ ,
\end{align}
where $S_{\text{BH}} = \mathcal{A}/4G$. 

It is important to pause and provide the necessary understanding, from the EFT standpoint, of the various quantities in the above equation. First, the quantity $\mathcal{A}_{\text{QG}}$ is intrinsically tied to the UV completion of quantum gravity. The constant $c^r_3(\mu)$,  and thus $\mathcal{A}_{\text{QG}}$, are in principle determined in either one of two ways. In the presence of a {\em full} theory of quantum gravity, one can extract $\mathcal{A}_{\text{QG}}$ by matching the partition function, eq.~\eqref{poartfun}, onto its counterpart in the full theory. Alternatively, $c^r_3(\mu)$,  and thus $\mathcal{A}_{\text{QG}}$, could be determined empirically at some infrared scale. Second, the quantity $\Xi$ comprises the massless spectrum of quantum gravity, which is an input parameter for the effective theory without any a priori constraints.
\\ \\
\emph{Constraints on the UV and IR - }Now we derive the main results of our letter. The basic premise in our construction is to adopt the Bekenstein-Mukhanov quantization conjecture. The fundamental statistical content of Eq.~(\ref{logentropy}) will allow us to derive two remarkable constraints on both $\Xi$ and $\mathcal{A}_{\text{QG}}$. The main power of the EFT is that it pins down the exact fashion that the UV physics, represented by $\mathcal{A}_{\text{QG}}$, is tied to the IR physics, represented by $\Xi$. Given the universality of the EFT, the constraints we derive below comprise consistency relations on the structure of any potential UV completion of quantum gravity, given the quantization conjecture is materialized.

We start by considering an adequate generalization to Eq.~\eqref{BMquant}, as first given in \citep{Hod:2004di} 
\begin{equation}\label{generalizedArea}
\mathcal{A}_n = \bigg(\gamma_0 n + \gamma_1 n^{\delta} + \gamma_2 \ln{n} \bigg)\, l_{\text{P}}^2, \quad n = 1,2,3,.....
\end{equation}
where $\gamma_0,\gamma_1,\gamma_2$ and $(\delta \geq 0)$ are constants. The generalized form in Eq.~\eqref{generalizedArea} is motivated by the logarithmic corrections to the BH entropy \citep{Hod:2004di}, where $n$ is interpreted as a quantum number associated to the area eigenstates of spherically symmetric black holes. Indeed, black hole thermodynamics necessitates an underlying statistical mechanical counting of the micro states \cite{Strominger:1996sh}. In particular, the total number of micro-states accessible to the system (i.e. the degeneracy of eigenstate $n$) is simply given by the exponential of the entropy
\begin{align}\label{conditionsgeneral}
g_n &\equiv \exp{\left(S_{\text{bh}}(n)\right)} \ \ ,
\end{align}
where, by construction, $g_n$ has to be a natural number $\mathbb{N}$ to guarantee consistency of the underlying theory. \\
\indent Given the black hole entropy in Eq.~\eqref{logentropy}, we must ask which values of $\mathcal{A}_{\text{QG}}$ and $\Xi$ would satisfy the requirement that $g_n \in \mathbb{N}$ for each and every eigenstate $n$. First of all, let us plug Eq.~(\ref{logentropy}) into Eq.~(\ref{conditionsgeneral}). One finds
\begin{align}\label{microstates}
g_{n} &\equiv \exp{\left(\frac{\mathcal{A}_n}{4 l_{\text{P}}^2}\right)}  \exp{\left( \Xi\ln\frac{\mathcal{A}_n}{\mathcal{A}_{\text{QG}}}\right)} \in \mathbb{N}, \: \forall n \ ,
\end{align}
and thus each exponential in the above expression must {\em individually} be a natural number given any eigenstate\footnote{Looking at Eq.~\eqref{microstates}, it is true that one might take each exponential to be a natural number up to a multiplicative $n$-independent constant. More explicitly, $\exp{\left(\mathcal{A}_n/4 l_{\text{P}}^2\right)} = R_1 \cdot x$ and  $\exp{\left( \Xi\ln{\mathcal{A}_n/\mathcal{A}_{\text{QG}}}\right)} = R_2 \cdot 1/x$, where $(R_1, R_2) \in \mathbb{N}$. Without loss of generality, we take $x=1$.}. Second, in the first exponential we substitute for $\mathcal{A}_n$ using the quantization rule, Eq.~\eqref{generalizedArea}, and find the following constraints on the constants
\begin{align}\label{conditions1}
\gamma_0 &= 4 \ln k,  \quad \gamma_1= 4\ln k_1, \quad \gamma_2 = 4q_1 \ \ ,
\end{align}
where
\begin{align}
k\in \mathbb{N}/\{1\}, \quad k_1 \in \mathbb{N}, \quad (\delta, q_1) \in \mathbb{N}\cup \{0\} \ \ .
\end{align}
Lastly, we examine the second exponential in Eq.~\eqref{microstates} and deduce the powerful condition 
\begin{align}\label{condition2}
\left( \frac{\mathcal{A}_n}{\mathcal{A}_{\rm{QG}}} \right)^{ \Xi} &\in \;  \mathbb{N}  \ \ ,
\end{align}
which, upon using Eq.~(\ref{generalizedArea}), becomes
\begin{align}\label{condbasic}
\left( \frac{ (\gamma_0 n + \gamma_1 n^{\delta} + \gamma_2 \ln{n})l_{\text{P}}^2}{\mathcal{A}_{\text{QG}}}\right)^{\Xi} \in \mathbb{N}, \: \forall n \ \ .
\end{align}
We immediately notice that it is impossible to satisfy the above condition, for each and every $n$, unless $\gamma_2  = 0$. Hence, Eq.~(\ref{condbasic}) simplifies
\begin{align}\label{condsimple}
\left(   \gamma_0  n  \frac{ l_{\text{P}}^2}{\mathcal{A}_{\text{QG}}}\right)^{\Xi}\left(1 + n^{\delta-1}\frac{\gamma_1}{\gamma_0} \right)^{\Xi}  \in \mathbb{N}, \quad  \forall n \ \ .
\end{align}
This relation is quite constraining, because each factor in the above equation must {\em individually} be a natural number. First off, one finds
\begin{align}\label{conditions2}
\frac{\gamma_1}{\gamma_0} =  s, \quad s \in \mathbb{N} \cup \{0\}\ \  .
\end{align}
Given that $\gamma_0$ and $\gamma_1$ are both logarithms of natural numbers, {\em cf.} Eq.~\eqref{conditions1}, it might prove impossible to satisfy the constraint in Eq.~\eqref{conditions2}. Therefore, it seems natural to make the simple choice
\begin{align}
k_1 = 1 \rightarrow \gamma_1 = 0 \ \ .
\end{align}
In addition, regardless of the what $\gamma_0$ and $\gamma_1$ might turn out to be, Eq.~\eqref{condsimple} presents us with a pair of remarkable constraints. First, the exponent $\Xi$ in Eq.~\eqref{condsimple} must be a natural number
\begin{equation}\label{Constraintonfields}
\frac{1}{180} \left(2N_S + \frac72 N_F - 26 N_V + 424 \right) = \bold{\mathscr{L}} \ \ ,
\end{equation}
where
\begin{align}
\bold{\mathscr{L}} \in \mathbb{N} \ \ .
\end{align}
Secondly, we have a constraint on the scale $\mathcal{A}_{\text{QG}}$ 
\begin{equation}\label{constraintonscale}
 \frac{\mathcal{A}_{\rm{QG}}}{l_{\text{P}}^2} = \frac{\gamma_{0}}{\bold{\mathscr{M}}^{1/\bold{\mathscr{L}}}} \ \ ,
\end{equation}
where another natural number emerges
\begin{align}
\bold{\mathscr{M}} \in \mathbb{N} \ \ .
\end{align}
The above two relations comprise the main results of this letter. The apparent simplicity of Eqs.~(\ref{Constraintonfields}-\ref{constraintonscale}) is quite striking and offer a prime example the power of the EFT framework. We stress that the validity of these constraints indeed relies on the Bekenstein-Mukhanov quantization conjecture.
\\ \\
\emph{Quantization rule for $c_3$ - }We can further process Eq.~(\ref{constraintonscale}) to derive a quantization rule for the Wilson coefficient, $c^r_3$, which is practically the directly measurable quantity in the low-energy effective action. Fixing a scale, say $\mu_\star$, in Eq.~(\ref{dimtrans})
\begin{equation}\label{constraintonc3}
c_3^r(\mu_*) = - \frac{\Xi}{64\pi^2} \ln{ (\mu_*^2 \mathcal{A}_{\rm{QG}})} \ \ ,
\end{equation}
we obtain
\begin{equation}\label{constraintc3_2}
c_3^r(\mu_\star) = \frac{1}{64\pi^2} \left[\ln \bold{\mathscr{M}} - \bold{\mathscr{L}}
 \ln \, \gamma_0 -  \bold{\mathscr{L}} \ln \left(\frac{\mu_\star^2}{8 \pi M_{\text{P}}^2}\right) \right] \ \ .
\end{equation} 
The above relation is quite non-trivial, because it forces the value of the Wilson coefficient, at any scale, to take on only very special values which must be expressible in terms of a single natural number $\bold{\mathscr{M}}$, once $\gamma_0$ and $\bold{\mathscr{L}}$ are fixed. This is, in essence, a quantization rule for the coefficient of the Kretschmann scalar in the low-energy limit of any UV model of quantum gravity. Although we can not make similar statements about the other constants in the action, i.e. those associated with the Ricci scalar and Ricci tensor, we conjecture that similar relations extend to the rest of the constants in Eq.~(\ref{localeff}).
 \\ \\
\emph{Field Counting -} Another interesting feature of Eq.~(\ref{constraintc3_2}) is that the size of $c_3^r$ depends crucially on IR data, given by the number of massless (light) fields in the theory. This furnishes a direct mixing between the UV and IR, which is completely unexpected from the EFT perspective. One could say that black hole thermodynamics provides a portal linking the IR to the UV. This raises the question: What {\em counts} as a light field in the vicinity of a black hole? This is a question where the EFT provides an answer.

We note that there are two mass scales in our partition function: the Planck mass, ${M_\text{P}}$, and the mass of the black hole, $M_{bh}$. For a field of mass $m$ to be considered light and enter $\Xi$, the condition  \cite{Donoghue:2015nba} is
\begin{equation}
\frac{1}{m^2} \int d^4x \sqrt{g} R_{\mu\nu\alpha\beta} \nabla^2 R^{\mu\nu\alpha\beta} \gg 1 \ \ ,
\end{equation} 
where $\nabla^2 = g^{\mu\nu} \nabla_{\mu}\nabla_{\nu}$. For Schwarzschild black hole this translates into
\begin{equation}\label{masscondition}
\frac{m}{M_\text{P}} \ll \frac{8\sqrt{3}\pi^2 M_\text{P}}{M_{bh}} \ \ .
\end{equation}
Thus, the heavier the black hole, the lighter the field has to be. A quick back-of-the-envelope calculation shows that for a solar mass black hole, only strictly massless particles would contribute to $\Xi$, if and only if we restrict ourselves to the standard model. Of course, in this case Eq.~(\ref{Constraintonfields}) hints at the presence of massless particles beyond the standard model.
\\ \\
\emph{Outlook -} To conclude, what can black holes teach us about the IR and UV? On the formal level, the results of this paper present profound glimpses into the the IR structure of any potential UV completion of quantum gravity. Given that any consistent model (or theory) of quantum gravity must possess a low-energy limit, Eqs.~(\ref{Constraintonfields}) and (\ref{constraintc3_2}) severely restrict the structure of the theory in the UV. Most notably, the massless spectrum in quantum gravity is intimately linked, through Eq.~\eqref{constraintc3_2}, to the classical limit of the quantum theory beyond general relativity. 

On the phenomenological level, it is hard to overstate the relevance of Eqs.~(\ref{Constraintonfields}) and Eq.~(\ref{constraintc3_2}). For starters, gravitational wave astronomy has lead to a surge in studying black hole solutions in theories beyond general relativity, in particular, quadratic gravity \cite{Svarc:2018coe, Lu:2015cqa, Holdom:2016nek}. The quasi-normal modes of such black holes is surely a function of the Wilson coefficient $c_3$, see for example \cite{Zinhailo:2018ska}. The quantization condition on $c_3$, Eq.~(\ref{constraintc3_2}), would then furnish interesting relations among the possible values of the quasi-normal frequencies. If such exotic black holes exist, and are detected via gravitational wave observations, then one can certainly hope to test the Bekenstein-Mukhanov quantization conjecture. 

Furthermore, Eq.~(\ref{Constraintonfields}) provides a potential constraint on new models of particle physics that predict new massless or light states, such as dark photons, light axions, and extra (possibly sterile) neutrino flavors. For example, we see that the simplest way of satisfying Eq.~(\ref{Constraintonfields}) is to have $N_S = 1$, $N_F = 4$, $N_V = 10$, which corresponds to $\bold{\mathscr{L}}=1$. Thus, given this number set we can merely accommodate, on top of the standard model, a single dark photon, one super-light axion in addition to a specified number of neutrino flavors (which depends on whether a standard model neutrino turns out to be massless).

\emph{Acknowledgements -} The work of BKE is supported in part by the Science and Technology Facilities Council (grant number ST/P000819/1). SM is supported by a Chancellor's International Research Scholarship by the University of Sussex. SM would like to thank Mark Hindmarsh for useful discussions.

\end{document}